\documentclass[aps,prb,floatfix,twocolumn,groupedaddress]{revtex4}
\usepackage[pdftex]{graphicx}
\usepackage{amsmath,amssymb,subfigure,color,verbatim,array} %epstopdf
\usepackage{yllmath}

\begin{document}
\title{Proposal for interferometric detection of topological defects in modulated superfluids}
\author{Mason Swanson}
\author{Yen Lee Loh}
\author{Nandini Trivedi}
\affiliation{Department of Physics, The Ohio State University, 191 W Woodruff Avenue, Columbus, OH  43210}
\date{June 20, 2011}
\maketitle

%@@@@@@@@@@@@@@@@@@@@@@@@@@@@@@@@@@@@@@@@@@@@@@@@@@@@@@@@@@@@@@@@@@@@@@@@@@@@@
%\blue{Note: Section headings will be gone eventually}
%\mysection{Introduction}
%@@@@@@@@@@@@@@@@@@@@@@@@@@@@@@@@@@@@@@@@@@@@@@@@@@@@@@@@@@@@@@@@@@@@@@@@@@@@@
\textbf{
Attractive interactions between fermions can produce a superfluid ground state, in which  pairs of up and down spins swirl together in a coordinated, coherent dance.
How is this dance affected by an imbalance in the population of up and down fermions? Do the extra fermions stand on the sides, or do they disrupt the dance?
The most intriguing possibility is the formation of a modulated superfluid state, known as an LO phase, in which the excess fermions self-organize into domain walls where the pairing amplitude changes sign.  
Despite fifty years of theoretical and experimental work, there has so far been no direct observation of an LO phase.
Here we propose an experiment in which two fermion clouds, prepared with unequal population imbalances, are allowed to expand and interfere.  A zipper pattern in the interference fringes is unequivocal evidence of LO physics.
Furthermore, because the experiment is resolved in time and in two spatial directions, we expect an observable signature even at finite temperatures (when thermal fluctuations destroy long-range LO order averaged over time).
} 

\begin{figure}[htb!]
	\centering
	\includegraphics[width=\columnwidth]{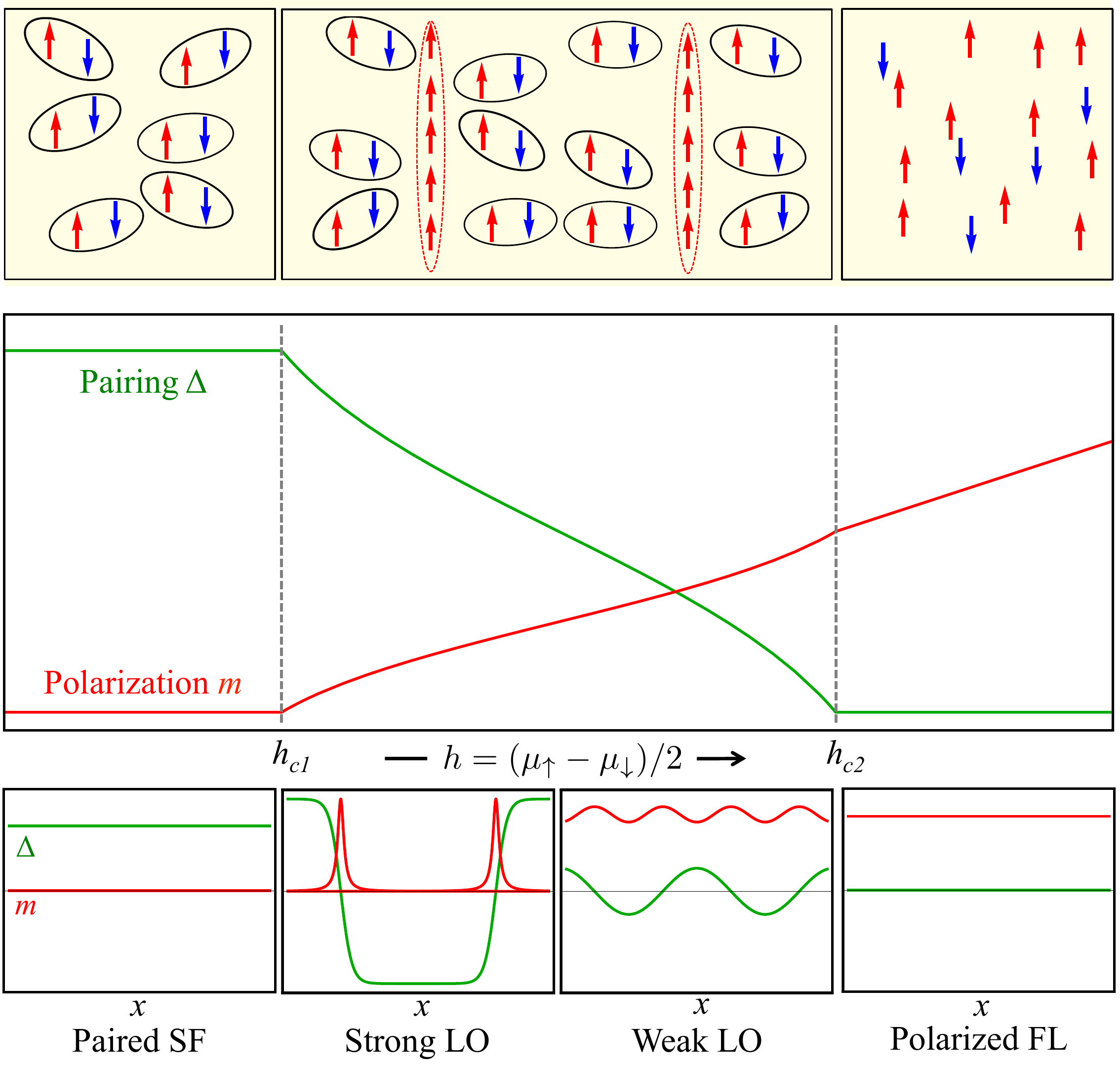}
	\caption{
	(Top) 
		Depictions of a fully paired superfluid,
		an LO state with excess fermions in domain walls,
		and a polarized Fermi liquid. 
	(Center) 
		%actually "RMS pairing amplitude".
		Pairing amplitude $\Delta$ and magnetization $m$ as a function of 
		the Zeeman field $h$, 
		which is the difference between the chemical potentials of up and down spins.
		The LO phase exists in a field range $h_{c1} < h < h_{c2}$.
	(Bottom) 
		Real space behavior of $\Delta(x)$ and $m(x)$ in each phase.
	\label{LOFigure}}
	\end{figure}

The study of modulated superfluidity has its roots in the context of a superconductor in a parallel magnetic field.
The inter-electron attraction favors a superfluid (SF) state consisting of pairs of up- and down-spin electrons,
whereas the field favors a polarized Fermi liquid (FL) state with a lower Zeeman energy.
The simplest theories predict a first-order SF-FL transition\cite{chandrasekhar1962,clogston1962}. 
However, the competition between pairing and polarization can produce far more subtle physics -- an intermediate Larkin-Ovchinnikov (LO) state~\cite{fulde1964,larkin1964,machida1984,burkhardt1994,yoshida:063601,loh2010}, as depicted in Fig.~1.
As the field is increased beyond $h_{c1}$ it forces excess fermions into the superfluid in the form of domain walls -- topological defects -- at which the order parameter changes sign between positive and negative values.  (This is analogous to how a perpendicular field forces vortices into a superconducting film.)
The wavelength of these modulations decreases with increasing field and ultimately the system gives way to a polarized Fermi liquid, shown in Fig. \ref{LOFigure}.

As a fascinating example of self-organized quantum matter, LO states have long been sought after in superconductors, in ultracold atomic gases~\cite{liao2010}, and even in neutron stars~\cite{alford2001}.
In the context of cold atoms, there have been proposals to detect LO states through
(a) modulations of the polarization in real space;
(b) peaks in the pair momentum distribution at the modulation wavevector;
(c) shadow features in the single-particle momentum distribution;
and 
(d) Andreev bound states in the density of states.
A recent experiment on an array of tubes found density profiles in agreement with Bethe ansatz calculations,~\cite{liao2010} which predict power-law LO correlations at zero temperature.
However, direct evidence of the modulations of the order parameter -- the defining feature of an LO state -- is still lacking.

%\cite{partridge2006,shin2006,liao2010,bloch2010}   %zhao:063605,

\begin{figure}[htb!]
	\centering
	\includegraphics[width=\columnwidth]{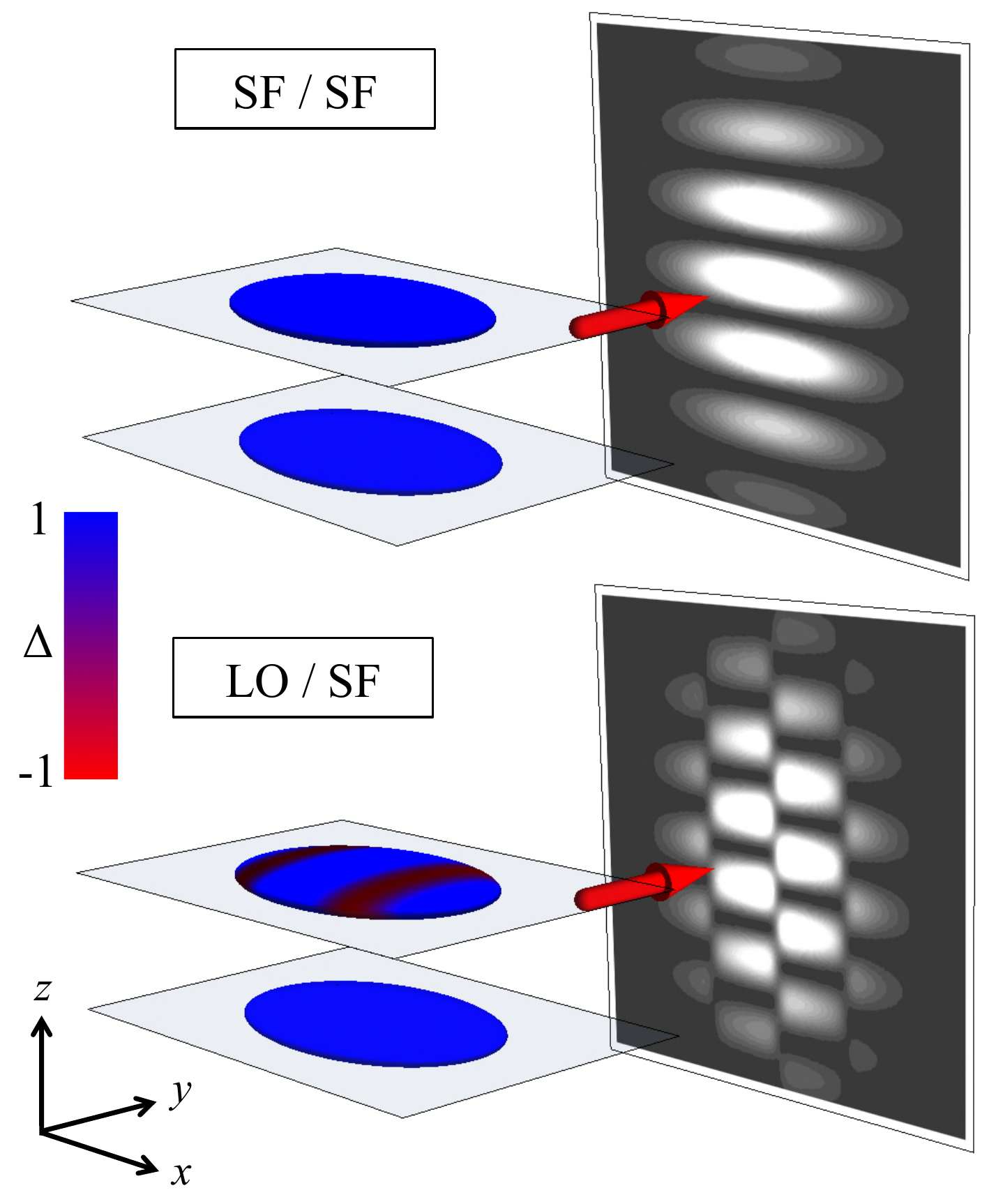}
	\caption{
		Proposed experiment and interferometric LO signature. An optical lattice is used to prepare two independent layers. These clouds are subsequently allowed to expand and interfere and are then imaged along one of the in-plane directions (here $y$). If both clouds are in the fully paired SF phase, the interference pattern is similar to the familiar double-slit experiment (top). However, if one of the layers contains an LO state (the change in sign of $\Delta(x)$ is indicated by color), the interference pattern contains sharp dislocations of the interference fringes (bottom).
	\label{experimentalIllustration}}
\end{figure}

\begin{figure}[t!]
	\centering
	\includegraphics[width=0.95\columnwidth]{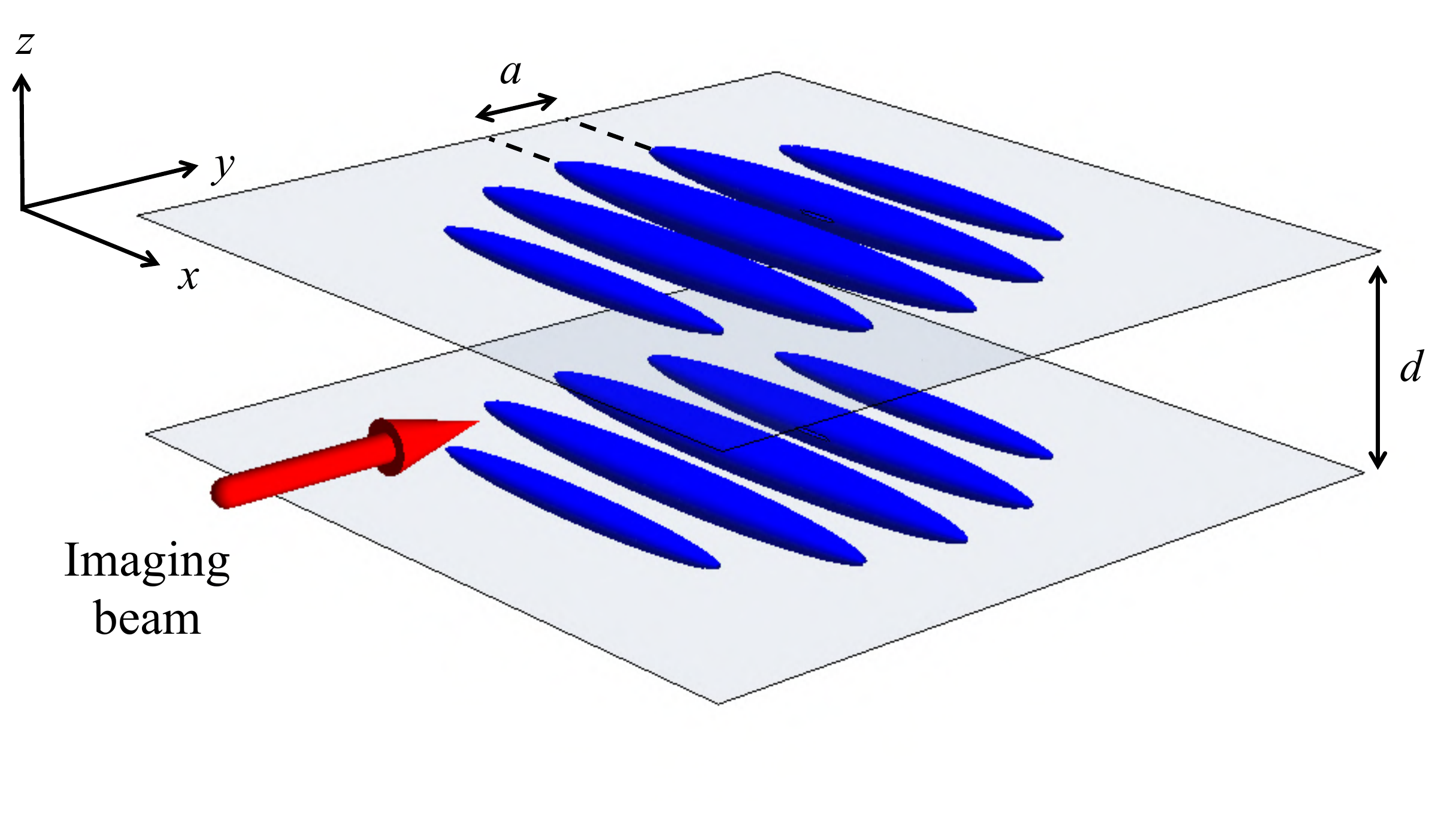}
	\caption{
		The fermion gas is confined in a harmonic trap.
		An optical lattice with a large spacing in the $z$-direction 
			is used to separate the gas into two independent quasi-2D layers. 
		A second optical lattice in the $y$-direction 
			cuts each layer into a series of weakly coupled tubes
			-- the optimal geometry for LO physics. 
		The trap and lattices are turned off abruptly,
			allowing the two layers to expand and interfere with one another.  
		Finally, the shadow of a probe laser in the $y$-direction
			gives the interference pattern projected onto the $x$-$z$ plane.
	\label{geometry}}
\end{figure}

%@@@@@@@@@@@@@@@@@@@@@@@@@@@@@@@@@@@@@@@@@@@@@@@@@@@@@@@@@@@@@@@@@@@@@@@@@@@@@

%@@@@@@@@@@@@@@@@@@@@@@@@@@@@@@@@@@@@@@@@@@@@@@@@@@@@@@@@@@@@@@@@@@@@@@@@@@@@@
 {\em Experimental Proposal:}
In this Letter, we propose an interference experiment in which two isolated superfluids expand into each other, as illustrated in Fig.~\ref{experimentalIllustration}.  
The ideal situation is shown in the lower panel of this figure, where one layer is a uniform fully-paired superfluid, which serves as a reference phase, and the other layer is a modulated (LO) superfluid.
The resulting interference pattern is directly sensitive to real-space modulations of the order parameter, and should provide an unequivocal signature of the elusive LO phase.

LO states have been predicted to exist in various situations \cite{loh2010,bulgac2008,cai2011}.  
The most likely of these to be realized in the near future is an array of tubes with a small intertube coupling.\cite{parish2007,zhao2008,rizzi2008}
This quasi-one-dimensional geometry provides good Fermi surface nesting at the LO wavevector $q_\text{LO} = k_{F\up} - k_{F\dn}$, together with Josephson coupling between the order parameter in adjacent tubes which is necessary to stabilize true long-range order.

Thus, we propose the following experiment.
Two species of fermions (referred to as $\up$ and $\dn$) are loaded into an optical trap.
The cloud is separated into two independent quasi-2D ``pancake'' layers using an optical lattice with a wide spacing in the vertical direction $z$
(created with two laser beams intersecting at a shallow angle).
\cite{hadzibabic2006}
The two layers are caused to have different population imbalances,
such that $n_\up^\text{top} \neq n_\dn^\text{top}$ but $n_\up^\text{bottom} = n_\dn^\text{bottom}$.
This may happen by chance due to natural number fluctuations;
alternatively, one can induce hyperfine transitions at the beat frequency between two laser beams, using a shallow-angle interference technique with appropriate phase shifts to address each layer separately.
An additional optical lattice is further used to create a 2D array of weakly-coupled tubes, conducive to the formation of an LO state.
The resulting geometry is shown in Fig.~\ref{geometry}.   

Once the gases are allowed to equilibrate for sufficient time, the interactions in the system are quickly ramped to the BEC side of the Feshbach resonance \cite{regal2004}
to ``freeze'' or ``project'' the pair wavefunction into a boson wavefunction, so that from this point onwards the pairs move as independent bosons (instead of disintegrating into fermions).
Then, the confining potentials are abruptly turned off.  As the clouds expand into one another, they interfere and form a 3D matter wave interference pattern. The projection of this interference pattern onto the $x$-$z$ plane can be measured by absorption imaging along the $y$ direction. Any LO phase modulation features will be captured in these projected interference patterns.

%@@@@@@@@@@@@@@@@@@@@@@@@@@@@@@@@@@@@@@@@@@@@@@@@@@@@@@@@@@@@@@@@@@@@@@@@@@@@@
%\mysection{Interference between coupled tubes}
%@@@@@@@@@@@@@@@@@@@@@@@@@@@@@@@@@@@@@@@@@@@@@@@@@@@@@@@@@@@@@@@@@@@@@@@@@@@@@

 \begin{figure}[h!]
	\centering
	\includegraphics[width=0.99\columnwidth]{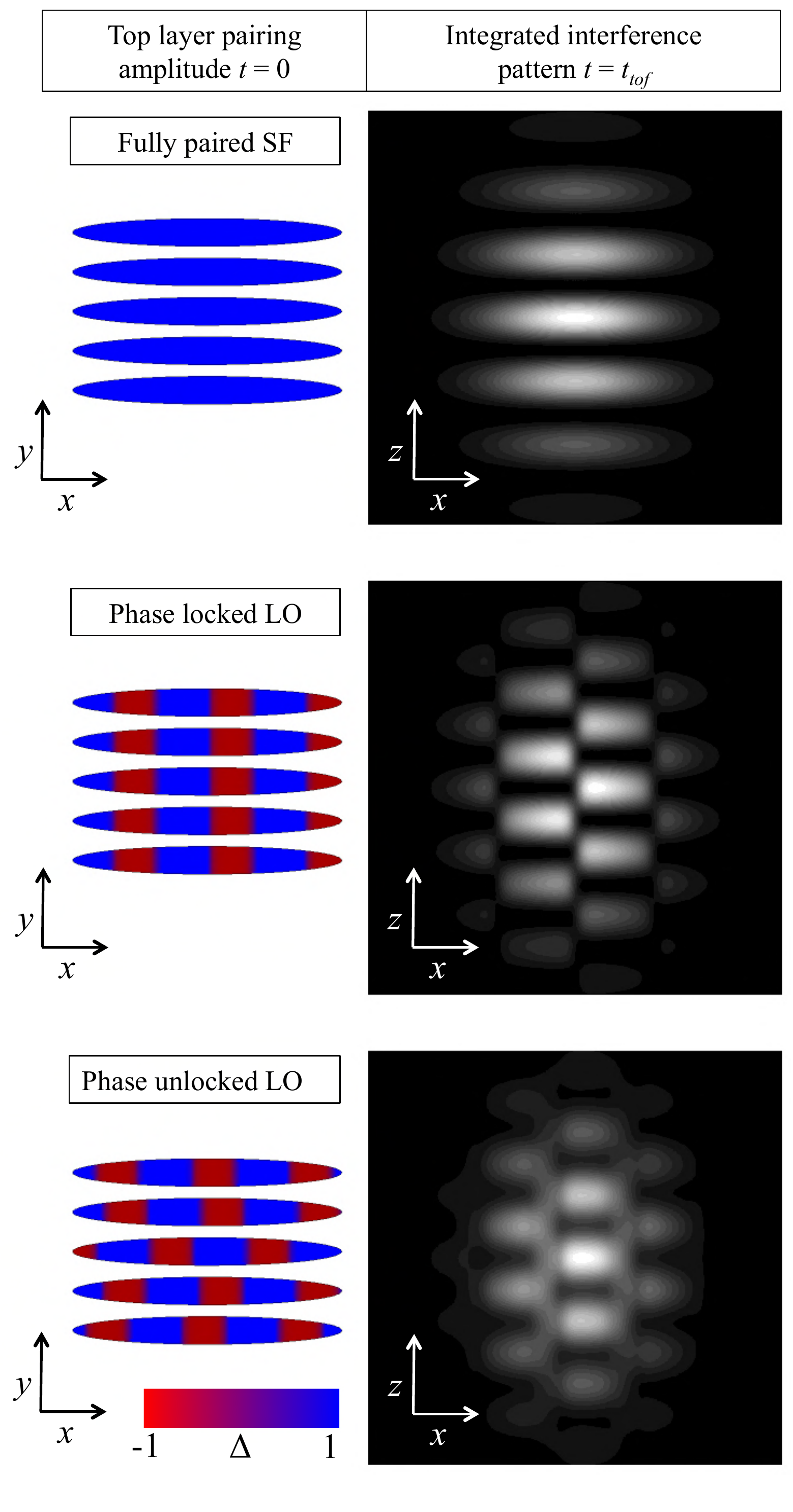}
	\caption{		Interference patterns for three different configurations: fully paired SF state (top), locked LO state (middle), and unlocked LO state (bottom). In each case we consider a $2 \times 5$ array of coupled tubes as shown in Fig. \ref{geometry}, using a fully paired SF as a reference phase in the bottom layer. The pairing amplitude $\Delta(x)$ in each of the five top layer tubes is shown to the left of the interference patternt. The locked LO states were taken to be $\Delta(x) = e^{-x^2/2 \sigma^2} \mbox{sn}(x/L | k)$, for $k \lesssim 1$ with period $L$, where
$\mbox{sn}(x | k)$ is the Jacobi sine function. In the unlocked case we added a random displacement of the domain walls  $\Delta(x) = e^{-x^2/2 \sigma^2} \mbox{sn}((x + \delta)/L | k)$ where $\delta \in [-L, L]$.  This interference pattern still contains signatures of the LO phase even when the domain wall locations fluctuate between tubes.}
%		Interference patterns for three different configurations: fully paired SF state (left), locked LO state (middle), and unlocked LO state (right). In each case we consider a $2\times 5$ array of coupled tubes as shown in Fig. \ref{geometry}, using a fully paired SF as a reference phase in the bottom layer. The pairing amplitude $\Delta(x)$ in each of the five upper layer tubes is shown to the right of the interference pattern with color indicating the respective tube for each amplitude $\Delta(x)$ (inset). The locked LO states were taken to be $\Delta(x) = e^{-x^2/2 \sigma^2} \mbox{sn}(x/L | k)$, for $k \lesssim 1$ with period $L$, where
%$\mbox{sn}(x | k)$ is the Jacobi sine function. In the unlocked case we added a random displacement of the domain walls  $\Delta(x) = e^{-x^2/2 \sigma^2} \mbox{sn}((x + \delta)/L | k)$ where $\delta \in [-L, L]$.  This interference pattern still contains signatures of the LO phase even when the domain wall locations fluctuate between tubes.}
	%[See previous versions for additional notes about parameters.]
	\label{illustrativeInterferenceComparison}
  \end{figure}

 {\em Interference between coupled tubes:}  
We now discuss analytically the salient features of the interference patterns of a 2D array of coupled tubes.  We begin by considering two layers, each containing $N$ coupled tubes, with separation $d$ in the $z$-direction.  After ramping up the interaction to produce a molecular BEC (of fermion pairs), the wavefunction is
  \begin{eqnarray} \nonumber
  \Delta(x,y,z) &=& \underbrace{e^{- (z - d/2)^2/2 \sigma_z^2} \sum_{n=1}^N \Delta^T_n(x) e^{-(y - an)^2/2 \sigma_y^2} }_{\mbox{top layer}} \\
                     &+& \underbrace{e^{- (z + d/2)^2/2 \sigma_z^2} \sum_{n=1}^N \Delta^B_n(x) e^{-(y - an)^2/2 \sigma_y^2}}_{\mbox{bottom layer}}
  \end{eqnarray}
where $a$ is the separation between in-plane tubes and $\sigma_y$ and $\sigma_z$ are the Gaussian confinements in the respective directions.  The wavefunction in the $n^{th}$ tube is denoted by $\Delta^T_n(x)$ in the top layer and by $\Delta^B_n(x)$ in the bottom layer.  
When the trap and lattices are switched off, the clouds expand predominantly in the tightly confined directions ($y$ and $z$).  After a suitable time of flight $t$, the wavefunction is effectively Fourier-transformed in the $y$ and $z$ directions.  That is,  the final wavefunction $\Delta(x,y,z,t)$ is approximately proportional to the initial momentum distribution in the $y$ and $z$ directions, i.e., $\Delta(x,y,z,t) \approx \Delta(x,k_y,k_z,t=0)$ where $y=tk_y/m$ and $z=tk_z/m$:
 \begin{eqnarray}  
  \Delta(x,k_y,k_z) &=&  \sigma_y \sigma_z e^{- z^2 \sigma_z^2/2}e^{-y^2 \sigma_y^2/2}  \\ \nonumber
  &\times& \left[e^{i k_z d/2} \sum_{n=1}^N \Delta^T_n(x) e^{ik_yan} \right.  \\ \nonumber
  &+& \left. e^{-i k_z d/2} \sum_{n=1}^N \Delta^B_n(x) e^{ik_yan} \right].
 \end{eqnarray}
 
The 3D density of the cloud, after expansion, is given by $I(x, k_y, k_z) \sim | \Delta(x, k_y, k_z) |^2$. 
The imaging process measures the integrated density along the $y$ direction,
 $I(x, k_z) \sim \int dk_y | \Delta(x, k_y, k_z) |^2$:
   \begin{eqnarray} \label{interferenceEq}
   I(x, k_z) &\sim& e^{-\sigma_z^2 z^2} \sum_{n=1}^N\sum_{m = 1}^N e^{-(n-m)^2 a^2/4 \sigma_y^2} \\ \nonumber
   &\times& \left[ \Delta^T_n(x) \Delta^T_m(x) +  \Delta^B_n(x)\Delta^B_m(x)  \right. \\ \nonumber
   && +  \left. 2 \Delta^T_n(x) \Delta^B_m(x) \cos k_z d \right].
   \label{InterferencePattern}
   \end{eqnarray}
   
We can consider the behavior of the above interference formula in its two limits: widely separated tubes ($a/\sigma_y \rightarrow \infty$) and overlapping tubes ($a/\sigma_y \rightarrow 0$).  In the overlapping limit, the total interference pattern reduces to that of two isolated 2D layers. In the limit of widely separated tubes, similar to the proposed experiment, the total interference pattern reduces to the sum of the interference patterns from adjacent tubes in the two layers.

The above analysis involves just two layers, which are necessary and sufficient for generating interference.  Introducing more layers would complicate the analysis, and reduce the visibility of the interference pattern; nevertheless, there would still be observable effects of the order of $1/\sqrt{N_\text{layers}}$.

% [Please see earlier versions for details on "Limits" of the intereference formula.]

\begin{figure}[htb!]
\centering
\includegraphics[width=0.95\columnwidth]{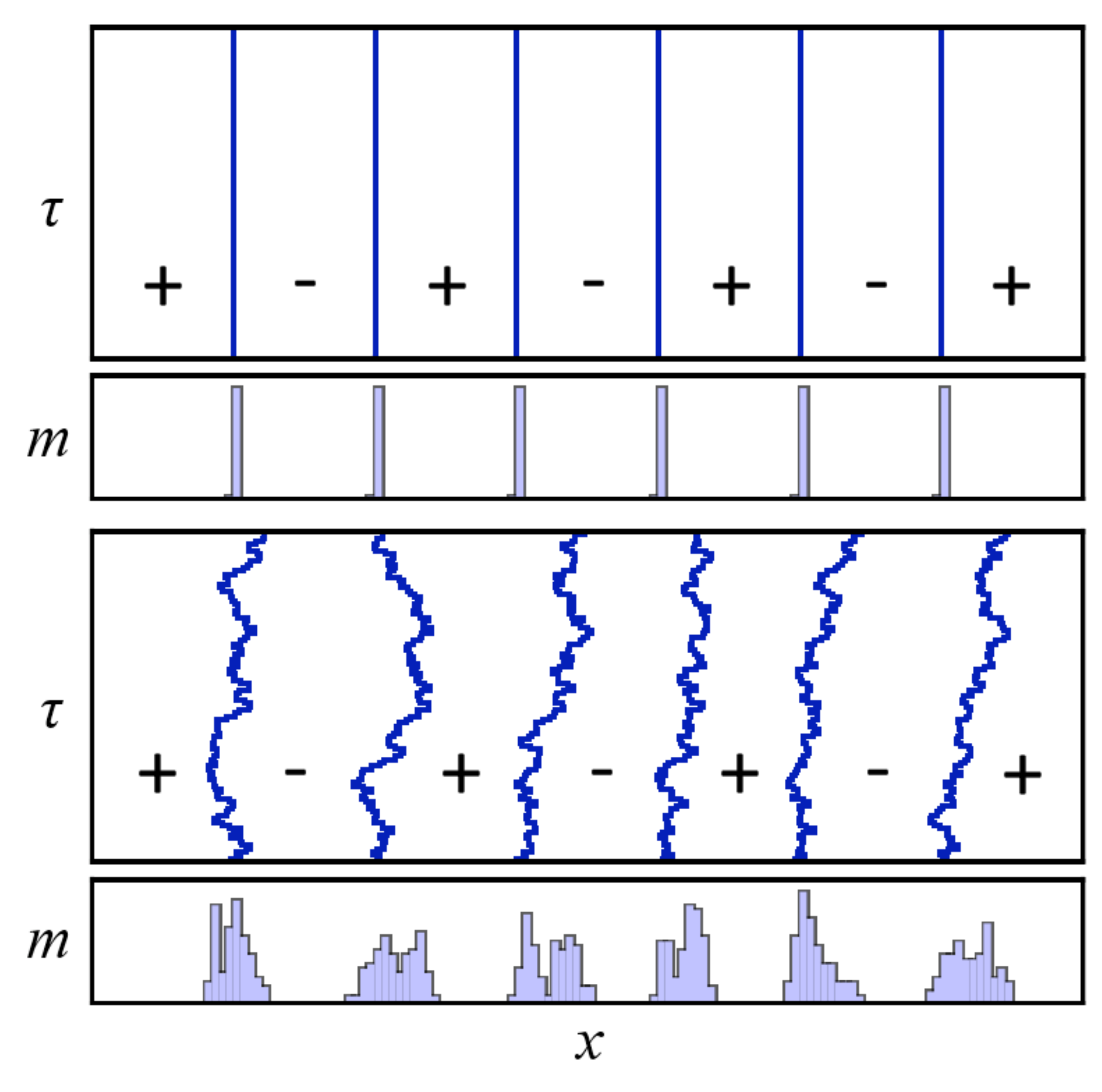}
\caption{
Illustration of imaginary time worldlines of LO domain walls and corresponding magnetizations with (bottom) and without (top) quantum fluctuations. Pluses and minuses correspond to the sign of the pairing amplitude $\Delta(x)$ in each region. In the bottom panel, the quantum fluctuations enforce a spatial profile for the magnetization that is less sharp, but which still clearly maintains the features of domain walls.
\label{quantumFluctuations}}
\end{figure}
   
Figure~\ref{illustrativeInterferenceComparison} illustrates the projected interference pattern, described by Eq.~\ref{InterferencePattern}, for three configurations of the upper layer:
a uniform SF; an LO phase with domain walls locked between tubes; and an LO phase with domain walls fluctuating between tubes. 
In each case we assume that the lower layer has been prepared in a uniform SF state.
We represent the LO pairing amplitude modulations
using Jacobi sine functions $\mathrm{sn}(x | k)$ \cite{lutchyn2010}
multiplied by Gaussian envelopes in the $x$, $y$, and $z$ directions.
The ``zipper'' pattern in the lower two panels is a clear signature of oscillations of the relative phase between the SF and LO layers, in contrast to the straight interference fringes in the top panel.

We have also obtained similar results from Bogoliubov-de Gennes (BdG) simulations.
The BdG simulations converge to ``locked LO'' states even when the intertube coupling is small.  Thermal phase fluctuations beyond BdG may be expected to produce domain wall fluctuations between tubes, as in the lowest panel of Fig.~\ref{illustrativeInterferenceComparison}, but even then a zipper-like pattern is still visible provided that the fluctuations are not too severe.

%@@@@@@@@@@@@@@@@@@@@@@@@@@@@@@@@@@@@@@@@@@@@@@@@@@@@@@@@@@@@@@@@@@@@@@@@@@@@@
%\mysection{Discussion}
%@@@@@@@@@@@@@@@@@@@@@@@@@@@@@@@@@@@@@@@@@@@@@@@@@@@@@@@@@@@@@@@@@@@@@@@@@@@@@

%{\color{blue}
%
%Need to discuss in what way interactions are turned off during the expansion process.  Ramping to BEC side?
%
%Do we want to make any comments on how the number of domain walls is important for feature detection in the interference pattern? What about the ideal dimensions of the geometry: the number of tubes, their size and separation, and the separation of the clouds?
%}

% Relation to Hulet P(q) proposal
 {\em Discussion:}
In an experiment previously proposed by the cold atom group at Rice, an LO state is prepared in a 2D array of tubes (with a spacing of about 532nm), the interaction is ramped to the BEC side, the trap is switched off, and the final density profile is a measure of the initial \emph{longitudinal} pair momentum distribution $P(q_x)$ (along the tube axis).  The function $P(q_x)$ is expected to have cusps at $\pm q_\text{LO}$, but these features tend to get washed out by temperature and spatial inhomogeneity, and their interpretation requires detailed quantitative analysis.
The key advantage of our proposal is that the clouds expand in the \emph{transverse} ($z$) direction, giving an interference pattern resolved \emph{along} the $x$-axis, which provides a much more direct way to probe the LO pairing amplitude modulations.

% Relation to Hadzibabic
The penetration of domain walls in an LO state resembles the penetration of vortices into a rotating BEC, and indeed, the zipper interference pattern we predict is similar to patterns that have been seen in experiments on vortices in 2D Bose systems \cite{hadzibabic2006}. 
Those experiments involved two pancake BECs separated by $3~\mathrm{\mu m}$.
%\blue{(To some extent they provide proof-of-principle for us.)}

%\blue{{\bf (Fluctuation effects:)} }
A remarkable feature of cold atom experiments is that control parameters (interaction, lattice depth, and trap depth) can be turned off very quickly, much faster than the typical timescale of domain wall movement.  This allows us to take a snapshot of the wavefunction (resolved in real time), in a way not possible in condensed matter experiments.  Thus, even above the critical temperature where thermal fluctuations destroy long-range order, it may \emph{still} be possible to detect LO physics in the form of ``temporary'' domain walls!

Quantum fluctuations of an LO state are more subtle.  Their effect can be thought of as diffusion of the domain walls in imaginary time $\tau$; measurements are necessarily averaged over $\tau$.  For isolated tubes, quantum fluctuations of the domain walls prevent long-range LO order, and there is only quasi-long-range order at zero temperature; hence, the interference pattern will be washed out.  This is why we recommend using sufficient coupling between the tubes to stabilize long-range order, so that the pairing amplitude modulations remain even after averaging over quantum fluctuations (see Fig.~\ref{quantumFluctuations}).

%\blue{{\bf (Conclusions:)} }
In a larger perspective, interferometric techniques have proven to be powerful methods to detect the relative phase of the pair wavefunction in 
cuprates,\cite{wollman1993} ruthenates, pnictides, 
and other unconventional superconductors.\cite{strand2009}
The interferometric signature we propose is different in that it involves a phase change in the center-of-mass modulations of the order parameter that occur across the domain walls
and not in the relative phase of the pair.

We gratefully acknowledge support from 
DARPA grant no. W911NF-08-1-0338 (YLL),
ARO W911NF-08-1-0338 (NT),
and the DARPA OLE Program (NT). 
MS acknowledges support from the NSF Graduate Research Fellowship Program.
We are grateful to Randy Hulet for useful discussions.

%{\it Correspondence and request for materials shoud be addressed to:} N. Trivedi
%(trivedi.15@osu.edu).  

%\bibliography{LO}
%\bibliographystyle{forprl}

\end{document}